\documentclass[a4paper,fleqn]{cas-sc}

\usepackage{amssymb}
\usepackage{subfig}
\usepackage{amsmath}
\usepackage{multirow}
\usepackage[dvipsnames]{xcolor}
\usepackage[authoryear]{natbib}
\usepackage{lineno}

\def\tsc#1{\csdef{#1}{\textsc{\lowercase{#1}}\xspace}}
\tsc{WGM}
\tsc{QE}

\begin{document}
\let\WriteBookmarks\relax
\def\floatpagepagefraction{1}
\def\textpagefraction{.001}

\shorttitle{Revisiting Dimorphos formation}
\shortauthors{Madeira \& Charnoz}
\title [mode = title]{Revisiting Dimorphos formation: A pyramidal regime perspective and application to Dinkinesh's satellite}

\author[1]{Gustavo Madeira}[
orcid=0000-0001-5138-230X,
]
\ead{madeira@ipgp.fr}

\affiliation[1]{organization={Université de Paris, Institut de Physique du Globe de Paris, CNRS},
            city={Paris},
            postcode={F-75005}, 
            country={France}}
\author[1]{Sebasti\'en Charnoz}

\cortext[1]{Corresponding author}

\nonumnote{}

\begin{abstract}
Dimorphos' oblate shape challenges formation models. Landslides on Didymos, induced by YORP effect, probably created a debris ring from which Dimorphos would have formed. Nonetheless, ring-derived satellites typically form with a prolate lemon shape. In light of the newest Dimorphos shape model, we revisit our previous work, \cite{MadeiraDimorphos}, and conducted new 1D simulations with material deposition in an extended region from 1.0 to 1.5 Didymos radii. An instantaneous landslide leads to a fast formation of a prolate Dimorphos directly from the ring. Now, if Didymos progressively deposits mass, Dimorphos grows through low-velocity impacts with large impactor-to-target mass ratio (pyramidal regime growth). Even during rapid, day-scale depositions, Dimorphos experiences one of these impacts, while for slower depositions, the satellite formation is primarily via pyramidal impacts. This process might reshape the satellite into an oblate shape \citep{Leleu2018} or even in a contact-binary shape, a scenario worthy of investigation that should be studied in the future with more suitable tools. Our conclusions can be applied to Dinkinesh's satellite, recently discovered by NASA's Lucy mission.
\end{abstract}

\begin{highlights}
\item Didymos deposits material into orbit via landslides.
\item An instantaneous single landslide typically results in a prolate lemon-shaped Dimorphos.
\item For slower depositions, Dimorphos grows via low-velocity impacts with a large impactor-to-target mass ratio.
\item For slower depositions, Dimorphos is expected to have a non-prolate shape. 
\item Low-velocity impacts with a large impactor-to-target mass ratio might also explain the shape of Dinkinesh's satellite
\end{highlights}

\begin{keywords}
Asteroids, dynamics \sep Debris disks \sep Planetary rings \sep Satellites, formation \sep Satellites, shapes
\end{keywords}

\maketitle


\defcitealias{MadeiraDimorphos}{M23}

\section{Introduction} \label{sec:intro}

Didymos' secondary, Dimorphos, was impacted in September 2022 by NASA's DART mission probe, responsible for the first astronomical-scale test of the kinetic deflection technology \citep{Thomas2023}. Located before the impact at $a_d=1183$~m from Didymos center, Dimorphos has an estimated mass and density of $M_d=4.3\times 10^{9}$~kg and $\rho_d=2400$~kg/m$^3$, respectively \citep{Naidu2020,Daly2022,Daly2023,Thomas2023}. As revealed by the data from DART mission, the object has a remarkably oblate shape, with an almost axisymmetric equatorial region \citep{Daly2023}, which proves to be an obstacle for the current models of Dimorphos formation. Didymos rotates near its spin limit and is believed to have experienced landslides and shape deformations in the past \citep{Yu2018,Zhang2021,Trogolo2023}. These events are believed to have given rise to a ring of debris from which Dimorphos is thought to have formed. Nevertheless, satellites formed directly from ring accretion are expected to exhibit a prolate shape \citep{Porco2007,Charnoz2010,MadeiraPhobos,MadeiraDimorphos}, as opposed to the oblate shape observed in Dimorphos \citep{Daly2023}. Here, we revisit the work of \citet[][hereafter M23]{MadeiraDimorphos} and explore, through 1D simulations, the conditions under which the satellite could have formed with a oblate (or even a contact-binary) shape. In Section~\ref{secrevisit}, we briefly describe the work of \citetalias{MadeiraDimorphos} and our new set of numerical simulations. Our results are addressed in Section~\ref{sec:dimorphos} and our conclusions drawn in Section~\ref{sec:discussion}.

\section{Revisiting the work of \cite{MadeiraDimorphos}} \label{secrevisit}

\citetalias{MadeiraDimorphos} propose Dimorphos' origin from a Didymos debris ring, originated from landslides due to Didymos spin-up driven by YORP effect \citep{Yu2018,Zhang2021,Trogolo2023}. Viscous ring spreading leads to the formation of proto-satellites at Didymos' Fluid Roche limit (FRL=$1.6926R_D$) and, ultimately, to the formation of Dimorphos \citep{Sugiura2021,Hyodo2022,MadeiraDimorphos}. \citetalias{MadeiraDimorphos} focuses on the viscous evolution of the ring and the formation of Dimorphos using the 1D hybrid code \texttt{HYDRORINGS} \citep{Salmon2010,Charnoz2010,Charnoz2011}. Two distinct scenarios are considered: one scenario with an initial ring of debris formed right at Didymos surface in a single instantaneous event, and another scenario in which the material is progressively deposited at Didymos' surface following the function for deposition of material:
\begin{equation}
\dot{M}(t)=\frac{M_T}{\tau}{\rm exp}\left(-\frac{t}{\tau}\right),
\end{equation}
where ${\rm M_T}$ is the total mass deposited in the ring, $t$ is the simulation time, and $\tau$ is the deposition timescale.

The mass of the largest satellite' formed primarily depends on ${\rm M_T}$ and 25\% of Didymos' mass \citep[${\rm M_D=5.6\times10^{11}}$~kg,][]{Daly2023} must be deposited in the disk to form a satellite with Dimorphos' estimated mass when the article was written. For $\tau\leq 1$~year, material spreading beyond the FRL coagulates into a proto-Dimorphos that grows in the ''continuous regime'' \citep{Crida2012}. In this regime, the newly-formed satellite grows by accumulating material directly from the ring. In other words, the ring's edge is inside the satellite's Hill sphere, so any material leaving the ring is accreted by the massive nearby satellite \cite[for a detailed description, see][]{Crida2012,Hyodo2015}. Due to ring torques, Dimorphos migrates outwards, while accreting  moonlets formed at the FRL that are in its gravitational reach \citep["discrete regime" growth,][]{Crida2012}. In this scenario, Dimorphos reaches its final mass in a few decades and is expected to be a prolate aggregate \citep{Porco2007,Charnoz2010}.

For $\tau\geq10^2$~yr, the picture is different: the satellite only reaches a few percent of Dimorphos' mass during the continuous regime. As the satellite moves away, the system evolves to a collection of similarly sized moonlets that eventually collide to form Dimorphos. This growth regime is defined as the ''pyramidal regime'' \citep{Crida2012}, and it takes thousands of years for Dimorphos to form. \cite{Leleu2018} shows that low-velocity impacts during the pyramidal regime may lead to the formation of oblate-shaped objects, similar to what is observed for Saturn moons, which could also be relevant for the oblate Dimorphos \citep{Daly2023}. 

However, there are some caveats regarding a slow formation of Dimorphos in the pyramidal regime. Numerical simulations indicate that landslides typically last only a few hours \citep{Sugiura2021,Hyodo2022}, a timescale considerably shorter than ours. Moreover, Didymos is subject to strong solar influence, which reduces the lifetime of material around the object \citep[see][]{Rossi2022,MadeiraVicinity}. Thus, a faster formation of Dimorphos is preferable.

Here, we present new simulations to explore Dimorphos' pyramidal regime growth more accurately. Our numerical model was improved using the secular perturbation theory, calculating first-order satellite interactions \citep{Brower1961,Dermott1986}. Secular interactions induce oscillations of orbital elements, and in each simulation step, we derive the secular terms of the satellites' disturbing function, solving them to determine the forced elements \citep[for details, see Chapter~7 of][]{Murray1999}. Additionally, we calculate the eccentricity kicks on a satellite due to the crossing of a first-order mean motion resonance \citep{Murray1999,Madeira2020}. With this, we can calculate the eccentricity of the satellites (at least in order of magnitude), which will allow us to crudely estimate the impact velocities.

To speed up simulations, we now start with a debris ring that occupies a broader region, rather than being confined to the primary's surface. Indeed \cite{Hyodo2022,Trogolo2023} have demonstrated that material is ejected over a much wider region than what was assumed in \citetalias{MadeiraDimorphos} and, based on their findings, we assume a uniform deposition of material from Didymos' surface \citep[$R_D=382$~m,][]{Daly2023} to $1.5R_D$. Didymos' tidal parameter is set as $k_2/Q=10^{-5}$. 

\section{Extended debris ring simulations and Dimorphos formation} \label{sec:dimorphos}
\begin{figure*}
\centering
\subfloat[]{\includegraphics[width=0.5\columnwidth,trim={0 0 0 0},clip]{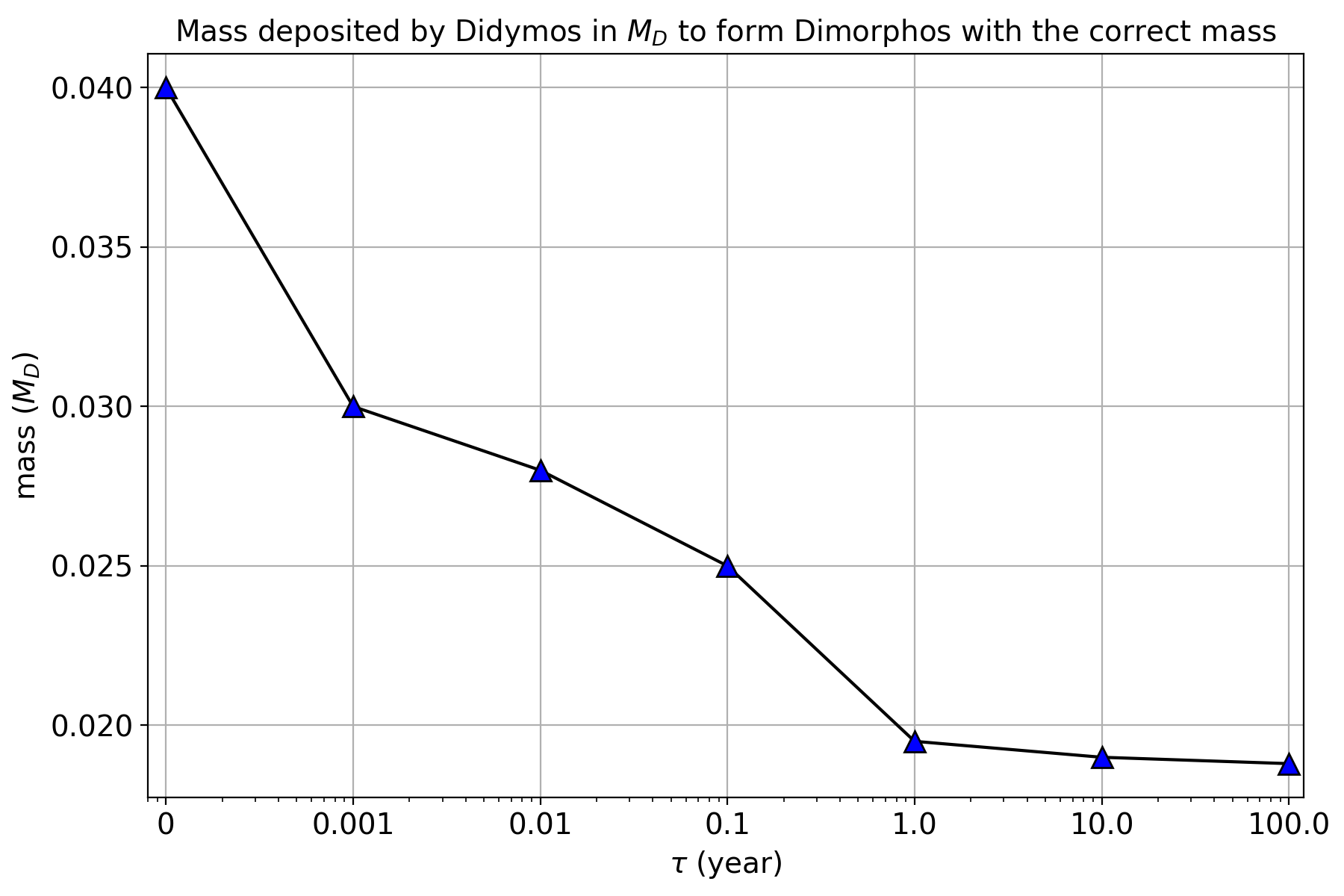}\label{figa}}
\subfloat[]{\includegraphics[width=0.5\columnwidth,trim={0 0 0 0},clip]{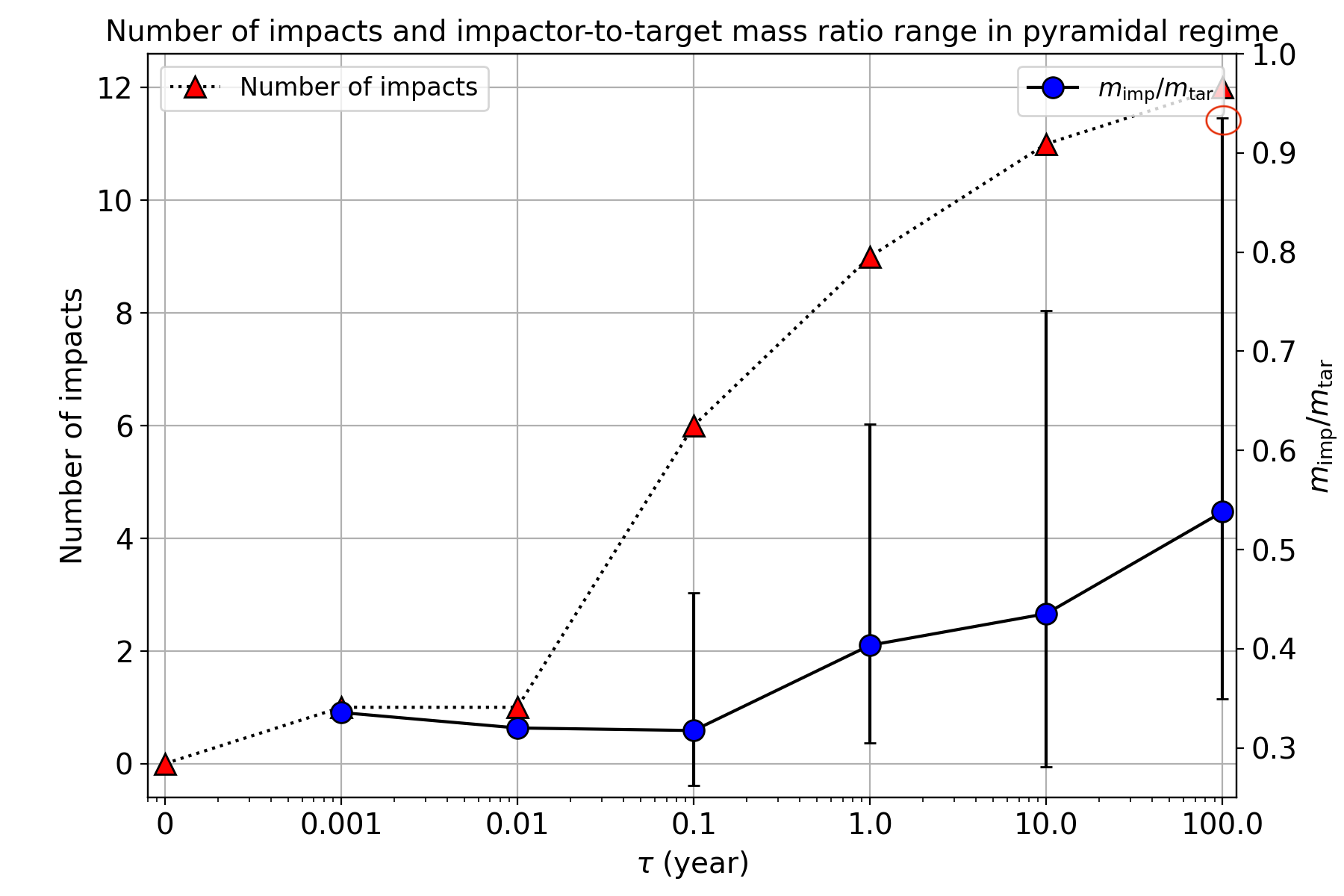}\label{figb}}\\
\subfloat[]{\includegraphics[width=0.5\columnwidth,trim={0 0 0 0},clip]{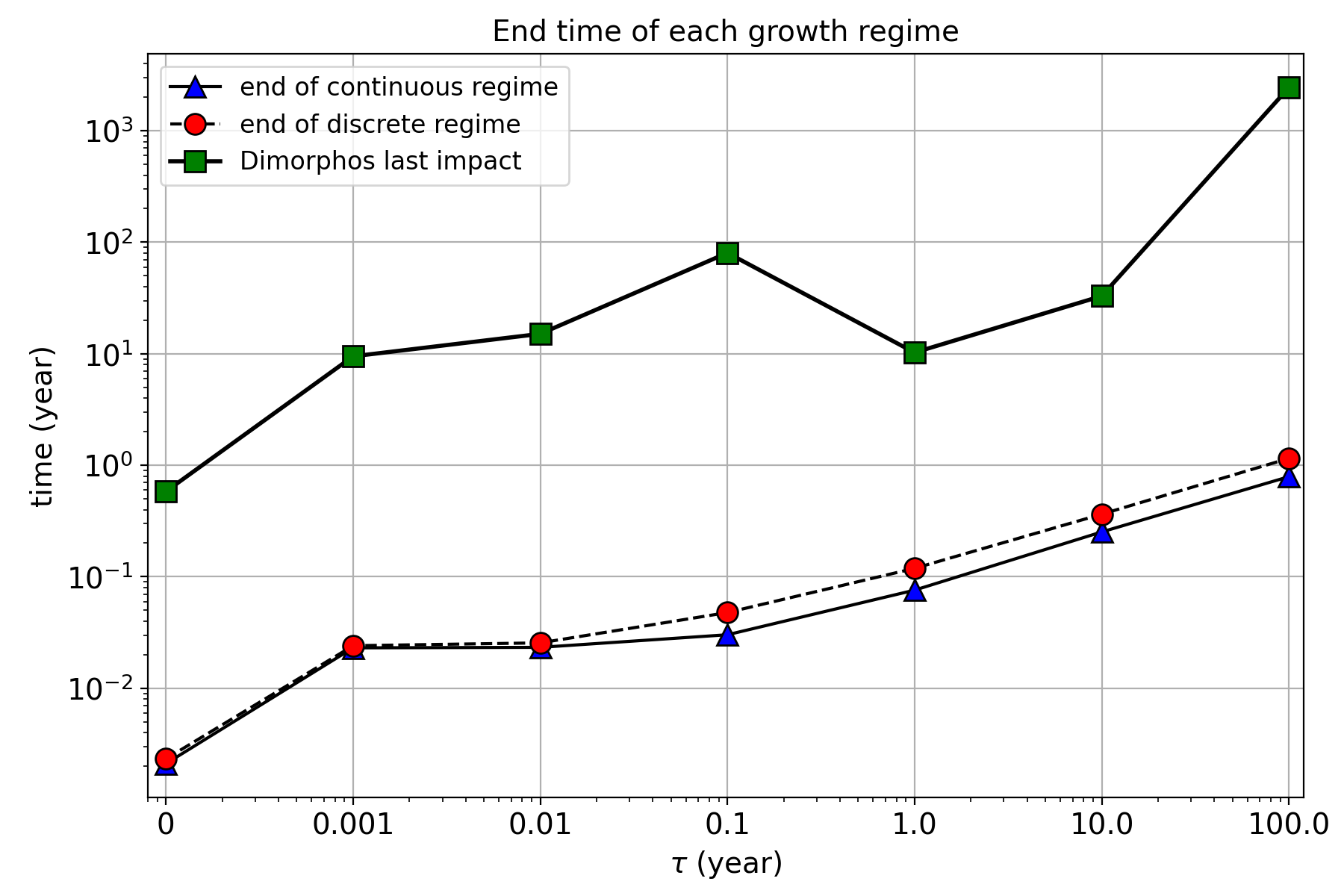}\label{figc}}
\subfloat[]{\includegraphics[width=0.5\columnwidth,trim={0 0 0 0},clip]{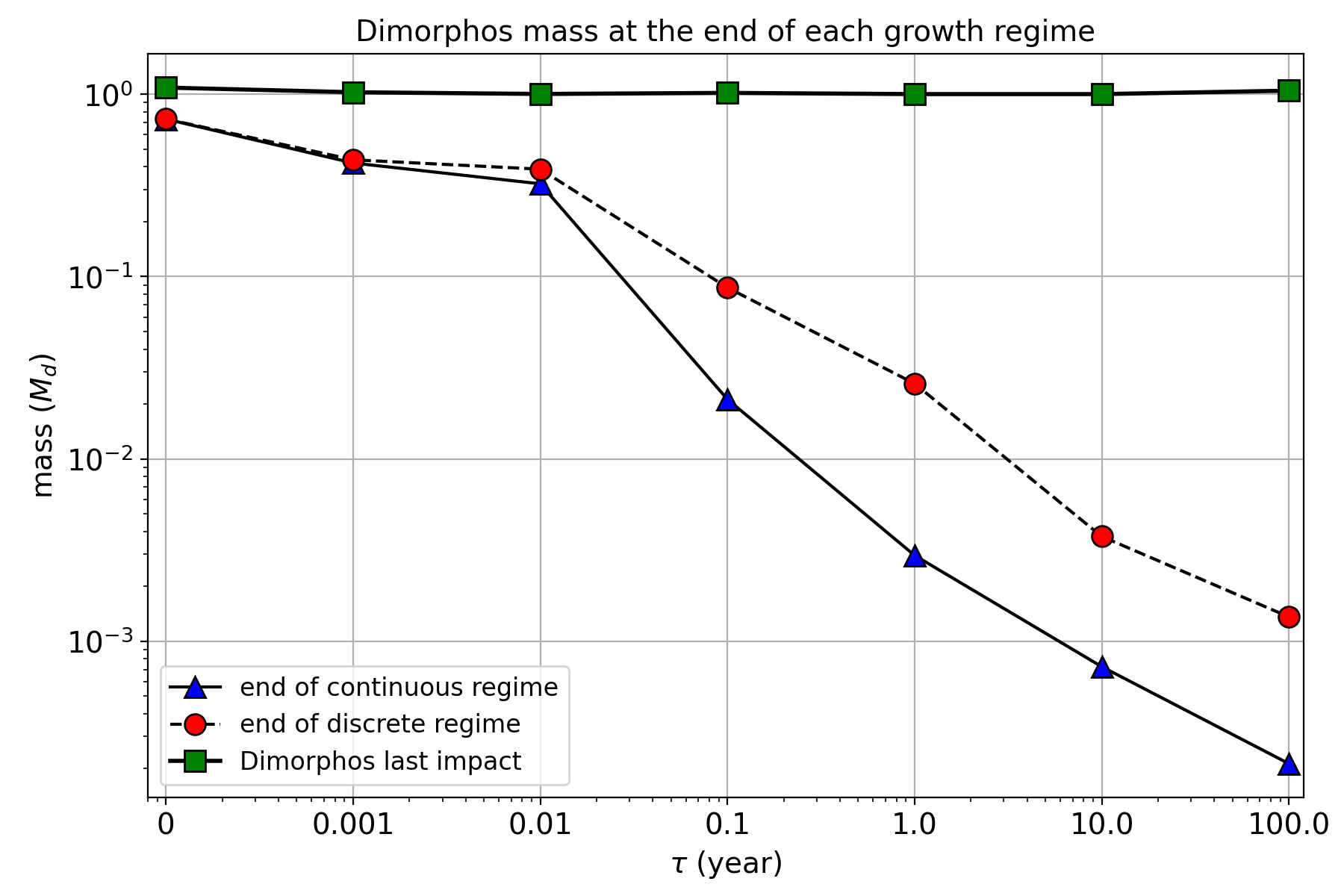}\label{figd}}
\caption{Summary of Dimorphos formation as a function of the deposition time scale $\tau$ with panels showing: a) total mass $M_T$ for Dimorphos formation. b) number of impacts in the pyramidal regime (left y-scale, dashed line) and the average impactor-to-target mass ratio m$_{\rm imp}$/m$_{\rm tar}$ (right y-scale, solid line) with error bars indicating the range of m$_{\rm imp}$/m$_{\rm tar}$. c) time and d) mass of Dimorphos at the end of the continuous regime, discrete regime, and last impact with Dimorphos.}
\label{fig}
\end{figure*}

We first consider the instantaneous formed ring case ($\tau=0$) and then explore scenarios where Didymos progressively supplies the ring over a deposition timescale $\tau$\footnote{This deposition timescale $\tau$ should not be confused with the disk lifetime $\tau_{\rm disk}$ in \cite{Crida2012}.} ($\tau=10^{-3}-10^2$ years). Our simulations are summarized in Figure~\ref{fig}. In the case of instantaneous deposition, Dimorphos can form from a ring with only $0.04~{\rm M_D}$ (Fig.~\ref{figa}), a much lower value than that reported in \citetalias{MadeiraDimorphos}, which assumes with a narrow, confined ring. This discrepancy is primarily due to the comparable angular momentum of the ring in both cases. For \citetalias{MadeiraDimorphos}, the total angular momentum brought into the system is $\sim2\times 10^{-12}~m^2/s$, while for the instantaneous formed extended ring the angular momentum is $3\times 10^{-12}~m^2/s$.

In the instantaneous case, proto-Dimorphos reaches $\sim0.74~{\rm M_d}$ in 21 hours ($\sim0.04$~year) in the continuous and discrete regimes (Fig.~\ref{figc},\ref{figd}). The satellite migrates outward until it reaches the 2:1 MMR position with the Roche limit (a$\sim$1027~m), which is the farthest distance reachable due to ring torques. At this location, Dimorphos achieves its final mass ${\rm M_d}$ in $\sim 6$~months through three impacts with impactor-to-target mass ratio of m$_{\rm imp}$/m$_{\rm tar}\sim5-15\%$. This differs from the typical pyramidal regime, as the impactors here are much smaller than proto-Dimorphos. In view of this, we define the limit of m$_{\rm imp}$/m$_{\rm tar}$=0.25 for the impact to be considered in the pyramidal regime. No impact meets this condition for $\tau=0$~year and therefore, a prolate Dimorphos is expected to form.

Now, let's explore scenarios in which Didymos progressively supplies the ring. As $\tau$ increases, the mass required for Dimorphos formation decreases. With higher $\tau$, the ring's viscous spreading rate slows down, causing less material to fall onto Didymos, thus increasing the efficiency of converting ring mass into satellites. However, this also results in the formation of smaller satellites at the Roche limit. Consequently, as $\tau$ increases, Dimorphos experiences reduced growth in the continuous and discrete regimes, while we observe an increase in the number of impacts in the pyramidal regime.

For $\tau=0.001$ and $0.01$~years, Dimorphos acquires $\sim0.4{\rm M_d}$ of mass in the continuous and discrete regimes, then experiencing just one pyramidal regime impact. This, however, might reshape it. With increasing $\tau$, Dimorphos mostly forms in the pyramidal regime, being obtained more collisions with higher mass ratio. These collisions occur at velocities of $\sim2-3$ mutual escape velocity and can lead to partial accretion or hit-and-run events, potentially leading to a non-prolate satellite \citep{Leleu2018}. This is particularly pronounced for $\tau=100$~yr, for which the maximum m$_{\rm imp}$/m$_{\rm tar}=0.94$ (red circle) and so, could give rise to a satellite with contact-binary shape. Impacts involving similar-sized objects at such velocities resemble the cases in which \cite{Leleu2018} obtain an oblate shape.

\section{Conclusion} \label{sec:discussion}
We revisit \cite{MadeiraDimorphos} and conduct new simulations with an extended disk and shorter mass-deposition timescales for Didymos. An instantaneous landslide on Didymos, distributing 4\% of its mass uniformly, would create Dimorphos in just a few hours, albeit likely with a prolate shape. However, if we assume a deposition event with a finite duration, the growth of Dimorphos directly from the ring is inhibited, and the satellite accumulates mass through low-velocity, high-mass-ratio collisions. Even in a deposition lasting a few days, Dimorphos undergoes one such impact, potentially altering its initial prolate shape. Longer duration events lead to almost complete growth in the pyramidal regime \citep{Crida2012}. The impacts associated to this regime may result in an oblate object, as observed by the DART mission. These conclusions certainly also apply to the recently discovered satellite of asteroid (152830) Dinkinesh by the Lucy mission. This moon is a contact binary, the result of a low-velocity encounter between two equally sized satellites, which defines the pyramidal regime.

\section*{Acknowledgments}
G.M. thanks to the Centre National d'Études Spatiales (CNES) and European Research Council (ERC). Numerical computations were performed on the S-CAPAD/DANTE platform, IPGP, France. We thank Aurélien Crida for the really nice comments that helped us to significantly improve the article.


\end{document}